\documentclass[manuscript]{aastex}
\usepackage{spr-astr-addons}
\usepackage{url}
\usepackage{hyperref}
\hypersetup{colorlinks,citecolor=blue,linkcolor=blue,urlcolor=blue}

\slugcomment{Not to appear in Nonlearned J., 45.}

\shorttitle{Magnetic field geometry of CB 34}
\shortauthors{Das et al.}

\begin{document}


\title{Magnetic field geometry of the large globule CB 34}


\author{A. Das}
\affil{Department of Physics, Assam University, Silchar 788011, India.}

\and
\author{H. S. Das}
\affil{Department of Physics, Assam University, Silchar 788011, India.}

\author{Biman J. Medhi}
\affil{Aryabhatta Research Institute of Observational Sciences, Manora Peak, Nainital 263129, India}

\and

\author{S. Wolf}
\affil{University of Kiel, Institute of Theoretical Physics and Astrophysics, Leibnizstr. 15, 24118 Kiel, Germany}




\begin{abstract}
We report the results of optical polarimetric observations of a Bok globule CB34 to study magnetic field structure on large scales ($10^5-10^6$ AU), which is combined with archival sub-mm observations to characterize the magnetic field structure of CB34 on small scales ($10^4 - 10^5$ AU). The optical polarization measurements indicate that the magnetic field  in the globule is constrained to a maximum radius of $10^5$AU around the core, out to densities not smaller than $10^4$cm$^{-3}$. Our study is mainly concentrated on two submillimeter cores C1 and C2 of CB34. The direction of magnetic field of  core C2 is found to  be nearly perpendicular to the CO outflow direction of the globule. The magnetic field of core C1 is almost aligned with the minor axis of the core which is typical for magnetically dominated star formation models. The mean value  of offset between the minor axis  of core C2 and  the outflow direction  is found  to be $14^\circ$ which suggests that the  direction of the outflow is almost  aligned with the minor  axis  of core C2. The magnetic  field  strength  in  the  plane-of-sky for cores C1 and C2 is  estimated  to  be  $\approx  34\mu$G   and  $\approx 70\mu$G.
\end{abstract}





\section{Introduction}
Bok globules are ideal sites for low mass star formation and the study of these objects can explore information about the earlier processes of star formation. The molecular clouds show structures on a variety of length scales, which are subdivided into \emph{clumps}, observed in CO. This clumps have characteristic masses $10^3 - 10^4$ M$_{\odot}$, radii 2--5 parsec, temperature of 10 K, mean number density of $H_2$ of $10^2 - 10^3$ cm$^{-3}$ and magnetic field $3 \times 10^{-5}$G. Higher-density cloud \emph{cores} are embedded in the clumps, observed in NH$_3$, CS, and other molecules. The typical sizes are 0.05--0.1pc, the temperature 10K, and the density $\approx 10^4 - 10^5$ cm$^{-3}$. The typical masses of core are about 1 to a few $M_{\odot}$, although a few range up to 1000 $M_{\odot}$. The clumps are surrounded by a low density \emph{envelope} which have a typical dimension of $\approx 10^4 - 10^6$ AU \citep{Bo2011}.

Magnetic fields play a significant role in collapse dynamics by mediating accretion, directing the outflows and collimating jets \citep{Li2004,Ga2009}. Fields may influence the shape of cloud fragments, the contraction timescale and the gas-dust coupling \citep{Mc2007}.
Magnetic field maps of the molecular clouds can be derived thanks to the effect of dust grain alignment \citep{La2007, Ho2014}. When the light from the background stars passes through a cloud of dust grains which are aligned with their minor axis parallel to the local magnetic fields of the cloud, it becomes linearly polarized due to selective extinction. Consequently, the magnetic field  can be detected through optical polarimetry in the low-density regions of molecular clouds. Complementary to this, the thermal reemission of aligned grains located in the high-density central region of
molecular clouds is linearly polarized perpendicular to the magnetic field direction and can thus be studied through submillimeter polarimetric measurement.
Numerous studies have been made on molecular clouds through optical and submillimeter polarimetry \citep{Ka1995,Wo2003,Al2008,Wa2009,Fr2010,Pa2012,Ch2014,Be2014,Ch2016}.

 Magnetic fields are believed to play a major role in the launching and collimation mechanisms of CO outflows, so it would be interesting to study the possible relationship between the observed magnetic field directions and outflows. \cite{Me2004} studied the alignment of T-Tauri stars with the local magnetic field which revealed a possible link between the strength of the Classical T-Tauri Star (CTTS) jets and their orientation with respect to the magnetic field. They concluded that the CTTSs with jets align to the magnetic field, but as a whole sample (i.e., both CTTSs with and without jets), the population is randomly oriented with respect to the magnetic field. \cite{Cu2007} presented and discussed polarization maps for 16 star-forming regions, obtained with the Submillimetre Common-User Bolometer Array (SCUBA) array on the James Clerk Maxwell Telescope (JCMT) at 850$\mu m$. They found no relation between mean field direction and outflow direction for the whole sample, although some alignments were noticed. \cite{Hu2014} found that the outflows are randomly aligned with B-fields, but sources with low polarization fractions showed hints of outflows being preferentially perpendicular to small-scale B-fields. \cite{Be2014} reported a magnetic field orientation parallel to the CO outflow in cloud CB54, while cloud B335 shows a change in the magnetic field orientation toward the outflow axis from the inner core to the envelope regions. In CB68, they found a magnetic field orientation almost perpendicular to the CO outflow. Recently, \cite{So2015} studied relative  orientation between magnetic field, minor axis of the cores and outflows for five dense cores IRAM 04191, L1521F, L328, L673-7 and L1014. They found that the outflows in three cores showed the alignment with the envelope magnetic field.

CB34 is a unique globule which is more massive than other globules and is situated at a distance of $\sim$ 1.5 kpc \citep{La2010}. The description of CB34 is presented in Table \ref{1}. This globule has three dense cores \citep{La1997,Hu2000}, and is also associated with many young stars which are believed to be formed from this globule \citep{Al1997}. The chemical age of CB34 was estimated to be more than $10^5$ yr \citep{Sc1998}. This young age is supported by the presence of a pre-main-sequence star, CB34FU, with an age of $\sim 10^6$ yr \citep{Al1997}. Several collimated outflows including a water maser \citep{Go2006} suggest the active ongoing star formation in CB34. A bipolar CO outflow from cool gas \citep{Yu1994a}, optical knots of atomic emission lines from radiative shocks (Herbid-Haro objects), and $H_2$ infrared jets \citep{Yu1994b,Kh2002} have been reported so far. The position angle of the bipolar CO outflow is $-15^\circ$ whose positions of the center of the outflows $\Delta \alpha$ and $\Delta \delta$ are given by $0.2'$ and $-0.4'$ (offset from the map center of CB34) \citep{Yu1994a}. The molecular environment of the young cluster of Class 0 YSOs located in the CB 34 was studied by \citet{Co2003} through a multiline millimeter survey. They reported that the present star formation activity is centralized in the three main clumps that have sizes of $\sim 0.25$ pc. \citet{Co2003} further found that either CB 34 is rotating, or that different parts of it are associated with different velocities.

 In contrast to previous studies, CB34 is located close to the galactic plane.
The well-studied magnetic field in this region of the galaxy will allow us to constrain how close to the globule the environmental magnetic field is dominating and where the magnetic field of the Bok globule is preferentially dominated by its internal field.

\begin{table*}
 \centering
  \caption{Basic properties of CB 34. Position angle of the major axis ($\theta_{maj}$), right ascension (RA), declination (DEC), galactic longitude and latitude ($l, b$), position angle of the galactic plane ($\theta_{GP}$), distance ($D$), position angle of outflow ($\theta_{out}$) and source classification (SC).}
  \begin{tabular}{cccccccc}
  \hline

$\theta_{maj}$ 	&	RA(2000)	&	DEC(2000)	    &	($l, b$) 	        &	$\theta_{GP}$        & $D$$^\dag$    &	 $\theta_{out}$$^\ddag$  	  &	 SC	 \\
 ($^\circ$)	&	(h m s)	    &	($^\circ$ $'$ $''$)	&  ($^\circ$)	&  ($^\circ$)   & (parsec)	 &  ($^\circ$)  &	  	 \\
\hline
    	90	        & 05:47:02.4	&	21:00:10     &	$(186.94, -3.83)$	    &	148.7	   & 1500        & $- 15$            &	class 0	 \\
																	
\hline
\label{1}
\end{tabular}

References: $^\dag$ \citet{La1997}, $^\ddag$ \citet{Yu1994a}\\

\end{table*}

\section{Observations}
The polarimetric observations have been performed using 1.04-metre Sampurnanand telescope of Aryabhatta Research Institute of observational sciencES (ARIES) near Nainital in India. The observations have been conducted on 12--13 March, 2013 and on 20 Feb, 2015. The ARIES Imaging Polarimeter (AIMPOL) has been used as a focal plane instrument, which consists of a  Wollaston prism and a rotatable half-wave plate (HWP). The function of Wollaston prism is to split the incident unpolarized beam into two rectangular polarized ordinary and extraordinary components whereas HWP changes the polarization state of the light wave. The imaging has been done using a CCD camera of 1024 $\times$ 1024 pixels.  This CCD corresponds to 1.73 arcsec per pixel with an effective field of view of about 8 arc minute  diameter on the sky. The gain  of the CCD is 11.98  e$^-$/ADU whereas the read-out noise is 7.0 e$^-$. The seeing radius is $\sim$ 4 arc sec. During all the phase of observations, broadband red filter ($\lambda$ = 630  nm, $\Delta$$\lambda$=120nm) has been used. The observational log is now presented in Table \ref{2}. The brief information about AIMPOL and data reduction procedures are discussed in \citet{Me2010} and \citet{Da2013}.

\begin{table*}
  \caption{Observation log.}
  \begin{tabular}{cccc}
  \hline

	Object	&		Name of 		&	Exposure 	&	Date	\\
	ID	&		Observatory   	&	time (sec)	        &    		\\
\hline
	CB34	 &  ARIES, Nainital	&	600	&	March 12--13, 2013 	 \\
																			
		     &  ARIES, Nainital  &	600	&	Feb 20, 2015	\\

\hline
\label{2}
\end{tabular}
\end{table*}

Standard stars with high and low degree of polarization have been also observed. The polarization details of the standard stars including offset are depicted in Table \ref{3}.

The linear polarization ($p$) and position angle of the polarization vector ($\theta$) in terms of the normalized stoke's parameters $q = Q/I$ and $u = U/I$ is given by

\begin{equation}
p = \sqrt{(q^2 + u^2)} ~~~~ \textrm{and} ~~~~ \theta = 0.5~\textrm{tan}^{-1}(u/q)
\label{1}
\end{equation}

\begin{table*}
\tabletypesize{\scriptsize}
\caption{Results of standard polarized and unpolarized stars at R-filter. p and $\theta$ from literature; p$_{obs}$ and $\theta$$_{obs}$ from observations.
Since the zero position of Half Wave Plate (HWP) is not systematically aligned with the north–south direction, offset angle is calculated using the relation: $\theta _0$ = ($\theta - \theta_{obs}$).}
\begin{tabular}{cccccccc}
\tableline
       UT date & Star & p (\%) & $\theta$ ($^\circ$)   & p$_{obs} (\%)$ & $\theta$$_{obs}$ ($^\circ$) & $\theta _0$ & Ref. for $p$ \& $\theta$\\
 \hline
March 12, 2013 & HD251204  & 4.79$\pm 0.4$  & 152.9              & 4.79$\pm 0.30$ & 155.7$\pm 1.8$  & --2.8$^\circ$ & $\dag$ \\
               & HD65583   & 0.01$\pm$0.02  & 144.7$\pm$30.0     & 0.15$\pm 0.20$ & 146.3$\pm$42 & --1.6$^\circ$ & $\P$ \\
               &           &               &              &       &               &  &\\
March 13, 2013 & HD251204  & 4.79$\pm 0.4$  & 152.9     & 4.86$\pm 0.32$ & 154.9$\pm 1.7$  & --2.0$^\circ$ & $\dag$\\
               & HD65583   & 0.01$\pm$0.02           & 144.7$\pm$30.0     & 0.12$\pm 0.18$ & 146$\pm$41 & --1.3$^\circ$ & $\P$  \\
      \hline
Feb 20, 2015   & HD155197  & 4.27$\pm 0.03$  & 102.9 $\pm$ 0.18 & 4.29$\pm 0.4$ & 117$\pm 2.6$ & --14.1$^\circ$ & $\S$ \\
               & HD109055  & 0.07$\pm$0.05            & 70.9    & 0.18$\pm 0.20$ & 85.2$\pm 29.8$  & --14.3$^\circ$  & $\ddag$\\
 \tableline
 \label{3}
\end{tabular}

$\dag$ http://www.sal.wisc.edu/HPOL/tgts/HD251204.html

$\ddag$ http://www.sal.wisc.edu/HPOL/tgts/HD109055.html

$\P$ \citet{Cl1981}

$\S$ \cite{Sc1992}

\end{table*}

\section{Geometry of magnetic field}
\subsection{Optical polarization}
\begin{figure*}
\hspace{1cm}
\includegraphics[width=20cm, height=10cm]{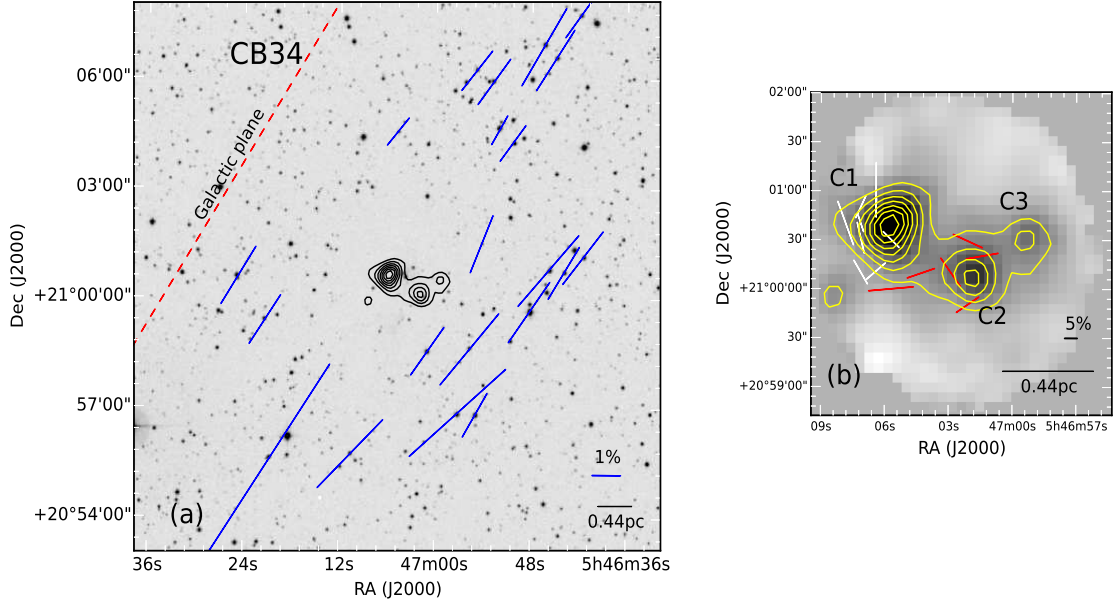}

\caption{(a) Optical polarization vectors (blue lines) are overplotted on a $\sim 15' \times 15'$ R-band DSS image of the field containing CB34. A vector with a polarization of 1\% is drawn for reference which represents that the length of the line segments are proportional to polarization percentage. The mean value of  optical polarization and position angle for     stars  projected in CB34  are $<p_{\textrm{opt}}>  = 2.14 \%$ and  $<\theta_{\textrm{opt}}>$ = 143.2$^{\circ}$ with a   standard   deviation    of   $\sigma_{p}$   =   0.84$\%$   and $\sigma_{\theta}$  =  7.8$^{\circ}$,  respectively.
The red dashed line represents the direction of the Galactic plane at the latitude of the cloud ($\theta_{\textrm{GP}} \sim 149^\circ$).  Contours correspond to SCUBA 850$\micron$ dust continuum emissions which range from 0.1 to 0.7 Jy beam$^{-1}$ in steps of 0.1 Jy beam$^{-1}$ \citep{Di2008}.   (b) Three sub-mm cores C1, C2 and C3 of CB34 are shown. Contours range from 0.1 to 0.7 Jy beam$^{-1}$ in steps of 0.1 Jy beam$^{-1}$. Sub millimeter polarization vectors (850$\micron$) are sampled on a $10''$ grid.  The sub-mm data, collected from the \citet{Ma2009} legacy data set, have been reanalyzed and presented in Table \ref{5}. Vectors are plotted where $I > 0$, $p/E_p > 2$ and $E_p < 6\%$.  White and red lines are vectors associated with C1 and C2 respectively. A vector with a polarization of 5\% is drawn for reference which represents that the length of the line segments are proportional to polarization percentage. The sub-mm polarization angles are rotated by 90 degrees to show the orientation of the magnetic field, for comparison with the optical polarization angles.  The mean value of sub-mm polarization  ($<p^{\textrm{sub}}>$)  and  the magnetic field position angle ($<\theta^{sub}_{\textrm{B}}>$) with corresponding standard deviations of  core  C1 are   given  by   $12.0 \pm 6.9\%$  and $65.4 \pm 69.3^\circ$. The values for C2 are given by $13.2 \pm 3.5\%$  and $88.3 \pm 33.5^\circ$. The intensity-weighted average magnetic field position angle ($<\theta^{sub}_{\textrm{B}}>_{wa}$) for cores C1 and C2 are given by $46.7\pm6.5^\circ$ and $90.4\pm12.3^\circ$.
}
\label{Fig1}
\end{figure*}

\begin{figure*}
\begin{center}
\vspace{-2 cm}
\hspace{-2 cm}
\includegraphics[width=30pc, height=20pc]{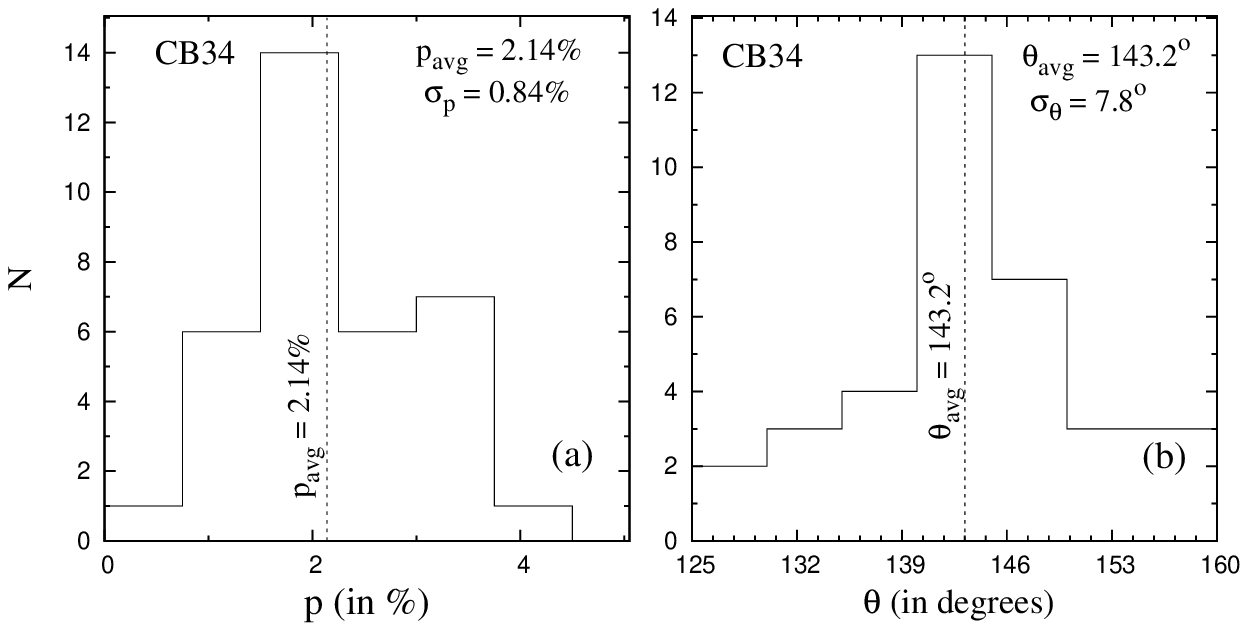}

\caption{Histogram shows number of star ($N$) versus  degree of linear polarization and  polarization position angle for CB 34. Dashed line in figure 2 (a) \& (b) represents the position of the mean value of polarization vectors and position angle respectively for the observed field stars of CB 34 (having $\textit{p}/\in_{p} \geq3$). We have excluded the high polarization star (\# 34 of Table-4). }
\label{Fig2}
\end{center}
\end{figure*}

The optical polarization percentage of all stars including position angles of polarization are calculated using equation (\ref{1}) and is presented in Table \ref{4}. In Fig.~\ref{Fig1}a, we have over plotted the polarization vectors of observed field stars on DSS R-band image of CB 34.  We considered 36 stars whose $p/E_p \ge 3$, where $E_p$ is the error in polarization. It is to be noted that $I/E_I > 15$, where $E_I$ is the error in intensity. The mean value of polarization of 36 stars is $p_{avg} = 2.30\%$ with a standard deviation $\sigma_p = 1.23\%$, also the mean value of position angle is $\theta_{avg} = 143.3^{\circ}$ with a standard deviation of  $\sigma_\theta = 7.7^{\circ}$.   It is also noticed from Table \ref{4} that the star \# 34 having $ p = 7.61 \pm 2.27$ (in \%) and $\theta = 147.0 \pm 8.6$ (in degrees), is highly polarized in comparison to $p_{avg}$ for rest of stars although it is almost aligned with rest of stars. If we exclude this star, the mean value of polarization and position angle for the rest 35 stars are $<p^{\textrm{opt}}> = 2.14 \%$ and $<\theta^{\textrm{opt}}>$ = 143.2$^{\circ}$ with a standard deviation of $\sigma_{p}$ = 0.84$\%$ and  $\sigma_{\theta}$ = 7.8$^{\circ}$ respectively. In Fig.~\ref{Fig2}, the distribution of degree of linear polarization and  polarization position angle versus number of stars for CB 34 are plotted. The histogram for CB34 shows narrow peak close to the mean polarization direction which is usually observed in the Galactic molecular clouds.

 Star \# 28, located in the southern region of projected image of CB34, is the nearest field star to CB34 which is traced at optical wavelength, having limiting distance $102''$ (corresponding to $\approx 1.5 \times 10^5$AU from the center of globule).
Constraints on the impact of the large-scale ambient magnetic field on the local magnetic field structure of the Bok globure can therefore be traced as close as to this distance.


\begin{table*}
 \centering
  \caption{Optical polarization data of CB34.}
  \begin{tabular}{cccccc}
  \hline

Star \#  & RA (2000)   &   DEC (2000)  &         $p$ (\%)   & $\theta$ ($^\circ$) & $p/E_p > 3$ \\

 \hline
1	&	05 46 28.35	&	21 04 30.61	&	0.62$\pm$0.20	&	143.16$\pm$9.24		& Y	\\
2	&	05 46 41.25	&	21 05 46.60	&	1.29$\pm$0.70	&	164.72$\pm$15.55	&	N\\
3	&	05 46 41.49	&	21 01 08.21	&	2.28$\pm$0.50	&	142.13$\pm$6.28	&	Y\\
4	&	05 46 42.03	&	21 07 42.99	&	1.77$\pm$0.56	&	144.11$\pm$9.06		&	Y\\
5	&	05 46 42.71	&	21 01 20.07	&	0.61$\pm$0.27	&	141.87$\pm$12.68	&	N\\
6	&	05 46 43.95	&	21 00 43.05	&	2.08$\pm$0.64	&	148.70$\pm$8.82		&	Y\\
7	&	05 46 44.99	&	21 03 24.20	&	1.36$\pm$1.19	&	156.53$\pm$25.07	&	N\\
8	&	05 46 45.01	&	21 06 33.83	&	2.45$\pm$0.62	&	147.37$\pm$7.25	    &	Y\\
9	&	05 46 45.45	&	21 04 47.15	&	1.34$\pm$0.93	&	142.06$\pm$19.88	&	N\\
10	&	05 46 45.78	&	21 00 46.72	&	3.21$\pm$0.25	&	139.02$\pm$2.23	    & Y	\\
11	&	05 46 46.51	&	21 06 55.07	&	3.06$\pm$0.25	&	149.67$\pm$2.34     & Y	 \\
12	&	05 46 47.47	&	20 46 49.97	&	3.15$\pm$1.00	&	126.83$\pm$9.10	    &	Y\\
13	&	05 46 47.65	&	21 09 49.24	&	1.86$\pm$0.27	&	150.44$\pm$4.16	    &	Y\\
14	&	05 46 48.11	&	21 05 58.47	&	0.72$\pm$0.30	&	154.20$\pm$11.94	&	N\\
15	&	05 46 48.30	&	20 45 55.81	&	3.19$\pm$1.01	&	143.90$\pm$9.07	&		Y\\
16	&	05 46 48.31	&	20 59 38.99	&	2.51$\pm$0.70	&	145.24$\pm$7.99	&		Y\\
17	&	05 46 49.91	&	21 10 42.38	&	1.23$\pm$0.36	&	158.92$\pm$8.39	&		Y\\
18	&	05 46 50.28	&	21 04 14.19	&	1.51$\pm$0.45	&	143.74$\pm$8.54	&		Y\\
19	&	05 46 50.70	&	21 10 02.47	&	1.28$\pm$0.27	&	139.77$\pm$6.04	&		Y\\
20	&	05 46 52.19	&	21 04 38.60	&	1.13$\pm$0.26	&	150.50$\pm$6.59	&		Y\\
21	&	05 46 52.92	&	21 05 54.93	&	1.92$\pm$0.60	&	143.97$\pm$8.95	&		Y\\
22	&	05 46 54.26	&	21 01 29.65	&	2.10$\pm$0.69	&	158.18$\pm$9.41	&	 Y	\\
23	&	05 46 54.93	&	20 56 50.00	&	1.72$\pm$0.52	&	150.23$\pm$8.66	&	 Y	\\
24	&	05 46 55.03	&	21 06 13.28	&	1.72$\pm$0.50	&	141.61$\pm$8.33	&		Y\\
25	&	05 46 55.39	&	20 44 24.87	&	2.31$\pm$0.75	&	156.97$\pm$9.30	&		Y\\
26	&	05 46 55.71	&	20 58 38.05	&	3.17$\pm$1.06	&	140.22$\pm$9.58	&	 Y	\\
27	&	05 46 57.22	&	20 56 51.73	&	4.48$\pm$0.72	&	131.83$\pm$4.60	&	 Y	\\
28	&	05 47 01.00	&	20 58 33.42	&	1.99$\pm$0.63	&	144.77$\pm$9.07	&		Y\\
29	&	05 47 03.51	&	21 08 57.66	&	2.03$\pm$0.53	&	140.57$\pm$7.48	&	Y\\
30	&	05 47 04.75	&	21 04 31.90	&	1.20$\pm$0.37	&	140.82$\pm$8.83	&	Y\\
31	&	05 47 04.93	&	20 47 46.00	&	2.04$\pm$0.61	&	146.54$\pm$8.57	&	Y\\
32	&	05 47 10.62	&	20 55 46.02	&	3.25$\pm$1.01	&	135.76$\pm$8.90	&	Y\\
33	&	05 47 17.64	&	20 47 42.44	&	1.67$\pm$0.52	&	141.14$\pm$8.92	&	Y\\
34	&	05 47 20.62	&	20 55 38.76	&	7.61$\pm$2.27	&	146.99$\pm$8.55	&	Y\\
35	&	05 47 21.34	&	20 59 24.98	&	1.99$\pm$0.51	&	146.89$\pm$7.34	&	Y\\
36	&	05 47 22.18	&	20 42 57.48	&	2.36$\pm$0.65	&	132.23$\pm$7.89	&	Y\\
37	&	05 47 24.81	&	21 00 36.62	&	2.30$\pm$0.76	&	148.40$\pm$9.47	&		Y\\
38	&	05 47 25.03	&	20 42 51.11	&	3.72$\pm$1.15	&	130.10$\pm$8.86	&		Y\\
39	&	05 47 29.10	&	20 48 29.04	&	1.12$\pm$0.33	&	136.21$\pm$8.44	&		Y\\
40	&	05 47 40.25	&	20 48 47.60	&	1.50$\pm$0.50	&	128.14$\pm$9.55	&		Y\\
41	&	05 47 41.30	&	20 49 57.56	&	1.23$\pm$0.38	&	142.08$\pm$8.85	&		Y\\

\hline
\end{tabular}
\label{4}

Note: ~~ Y: Yes, ~~~ N: No
\end{table*}


\subsection{Submillimeter polarization}
 The sub-mm polarization data of CB34 is taken from the \citet{Ma2009} legacy data set. The submillimeter observation of CB34 were made at the James Clerk Maxwell Telescope (JCMT), Mauna Kea, Hawaii, using the Submillimeter Common User Bolometer Array (SCUBA). The polarization data were sampled at a 10$''$ pixel grid in J2000 coordinates. \citet{Ma2009} selected only those measurements for which $I > 0$, $p/E_p > 2$ and $E_p < 4\%$. This criteria restricted the data points to only six (five data points are associated with core C1 and one data with core C2; ref. Fig. 31 of \cite{Ma2009}). To study the magnetic field properties of cores C2 and C3, we have reanalyzed the data and calculated polarization vector at each pixel and use the criteria $I > 0$, $p/E_p > 2$, and $E_p < 6\%$ (instead of $< 4\%$). It is to be noted that $I/E_I > 23$, where $E_I$ is the error in intensity. The polarization data is now presented in Table \ref{5}. We understand this way we are considering some low polarization signal.

\begin{figure*}
\begin{center}
\vspace{5 cm}
\hspace{-4 cm}
\includegraphics[width=30pc, height=20pc]{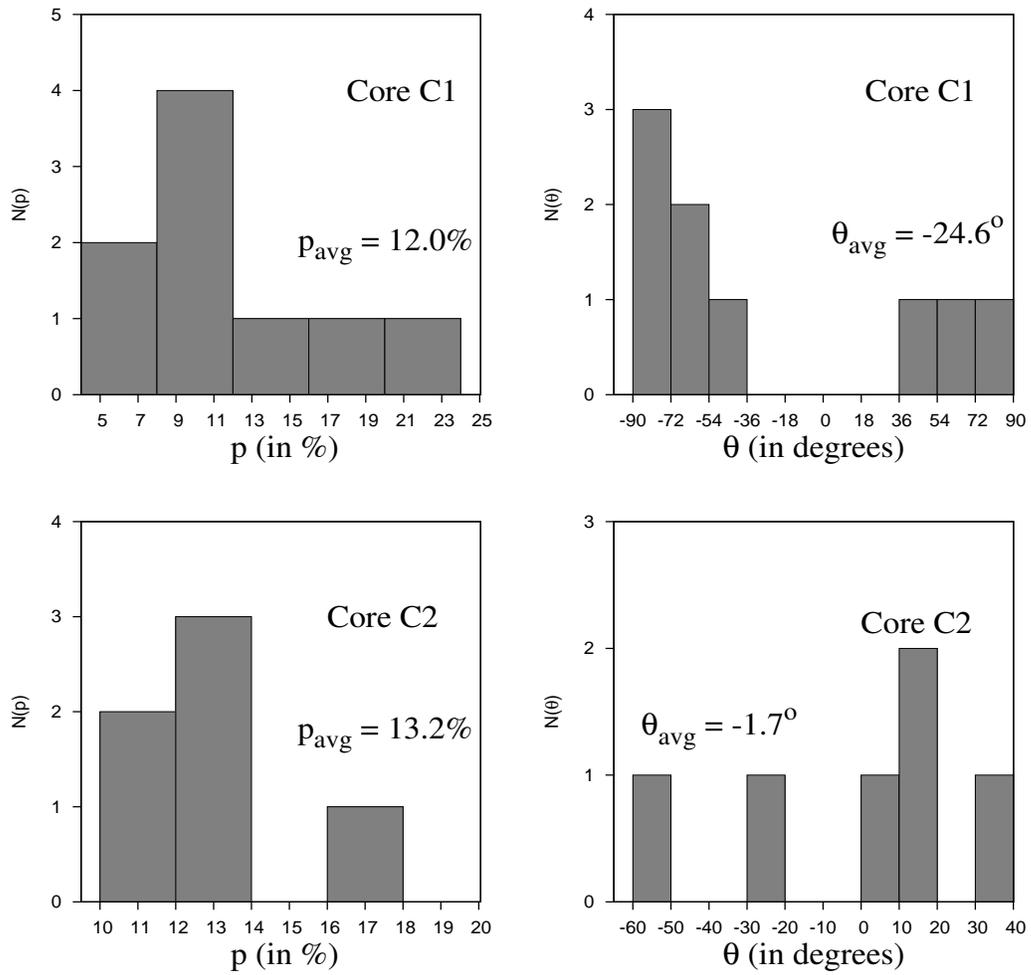}

\caption{Histogram shows the distribution of degree of linear polarization and  polarization position angle for two cores (C1 and C2) of CB 34.}
\label{Fig3}
\end{center}
\end{figure*}

It is to be noted that the sub-mm polarization angles have to be rotated by 90 degrees to show the orientation of the magnetic field, for comparison with the optical polarization angles \citep{Wo2003}. The magnetic field position angles ($\theta_B^{sub}$) are shown in the sixth column of Table \ref{5}. The sub-mm polarization map is now presented in Fig.~\ref{Fig1}b. This map shows almost randomly oriented magnetic field directions spread over the two brightest cores in the globule, and fields in these two cores appear to be unrelated to each other. It is further noticed that the   magnetic field vectors are associated with two cores (C1 and C2) only, not with core C3. So we will study the magnetic field geometry of two separate cores C1 and C2. In Fig.~\ref{Fig1}b, white and red lines are drawn to represent  magnetic field vectors of two cores C1 and C2 respectively. We assume that the alignment of these  field vectors are due to local magnetic field of individual core.  The mean polarization ($<p^{\textrm{sub}}>$) and mean position angle ($<\theta^{sub}>$) of core C1 along with standard deviations are given by $12.0 \pm 6.91\%$  and $-24.63 \pm 69.30^\circ$ (or $155.37 \pm 69.30^\circ$). The values for C2 are given by $13.18 \pm 3.46\%$  and $-1.72 \pm 33.52^\circ$ (or $178.28 \pm 33.52^\circ$). We find that the standard deviations of $<p^{\textrm{sub}}>$ and $<\theta^{sub}>$ for core C2  are less as compared to core C1. To study average magnetic field orientations, it would be good to perform a signal-to-noise-weighted angle average so that the average orientation is weighted in favor of the detections with the highest signal-to-noise ratio (SNR). The intensity-weighted average position angle\footnote{$\mathbf{<\theta^{sub}>_{wa} = \frac{\sum_{i=1}^n [\theta_i (1/E_{\theta_i}^2)]}{\sum_{i=1}^n [1/E_{\theta_i}^2]}}$}($<\theta^{sub}>_{wa}$) for cores C1 and C2 are given by $-43.3\pm6.5^\circ$ (or $46.7\pm6.5^\circ$) and $0.4\pm12.3^\circ$ (or $90.4\pm12.3^\circ$). Also, the weighted standard deviation of position angle ($\sigma^{wt}_{\theta}$) for cores C1 and C2 are estimated to be 51.9$^\circ$ and 25.5$^\circ$.
 In Fig.~\ref{Fig3}, the distribution of degree of linear polarization and  polarization position angle for two cores (C1 and C2) of CB 34 are plotted.

We now estimate core FWHM dimensions and position angle of the minor axis ($\theta_{min}$) of C1 and C2 from 850 $\micron$ map of CB34, which are given by ($\sim 36'' \times 33''$, $\sim 61^{\circ}\pm 5^{\circ}$)  and  ($\sim 35'' \times 32''$, $\sim -1^{\circ}\pm 5^{\circ}$).

The furthest sub-mm polarization vector (\#1 of core C1, see Table \ref{5}) of CB34 is located $85''$ ($\approx 1.3 \times 10^5$AU) away from the center of this globule. Thus the magnetic field at sub-mm wavelength can not be detected beyond $1.3 \times 10^5$AU. In the Southern region, the spatial gap between polarization vectors in the sub-mm and optical is $82''$ ($\approx 1.2 \times 10^5$AU) whereas this gap in the Western region is $120''$ ($\approx 1.8 \times 10^5$AU).

\begin{table*}
 \centering
  \caption{Submm polarization data of CB34 with $I > 0$, $p/E_p > 2$ and $E_p < 6\%$. The table is created from the CADC repository of the SCUBA Polarimeter Legacy Catalogue \citep{Ma2009} after reanalyzing the data set.   The first column represents the serial number of sub mm polarization vectors whereas the second and the third column represent pixel position of vectors in equatorial coordinate system. The fourth and fifth column are for degree of linear polarization and polarization angle of vectors. To compare with the optical polarization angles, the sub-mm polarization angles have been rotated by 90$^\circ$ to indicate the orientation of the magnetic field. The magnetic field position angles ($\theta_B^{sub}$) are shown in the sixth column. In the last column, the polarization vectors associated with cores C1 and C2 are shown.}
  \begin{tabular}{ccccccc}
  \hline

\#  & RA (2000)   &   DEC (2000)  &         $p$ (\%)   & PA  ($^\circ$) & $\theta_B^{sub}$ & Core \\
     &             &               &                    &                & (= PA + 90$^{\circ}$) & \\

 \hline
1	&	05	47	7.8	&	21	00	40.2	&	18.79	$\pm$	3.79	&	$-$70.78	$\pm$	5.91	&	19.22 &	 C1\\
2	&	05	47	7.1	&	21	00	10.2	&	11.04	$\pm$	4.28	&	$-$61.02	$\pm$	8.7	    &	28.98 &	 C1\\
3	&	05	47	7.1	&	21	00	31.2	&	12.93	$\pm$	5.2	    &	$-$77.43	$\pm$	6.36	&	12.57 &	 C1\\
4	&	05	47	7.1	&	21	00	40.2	&	07.30   $\pm$	2.25	&	$-$72.08	$\pm$	9.01	&	17.92 &	 C1\\
5	&	05	47	7.1	&	21	00	50.2	&	9.76	$\pm$	3.36	&	66.26	$\pm$	9.06	&	156.26 &	 C1\\
6	&	05	47	6.4	&	21	00	10.8	&	10.05	$\pm$	4.16	&	42.07	$\pm$	9.98	&	132.07 &	 C1\\
7	&	05	47	6.4	&	21	00	50.2	&	6.77	$\pm$	2.96	&	87.16	$\pm$	16.2	&	177.16 &	 C1\\
8	&	05	47	6.4	&	21	01	00.6    &	22.33	$\pm$	4.55	&	$-$89.89	$\pm$	11.31	&	0.11 &	 C1\\
9	&	05	47	5.7	&	21	00	00.0    &	17.61	$\pm$	4.58	&	5.43	$\pm$	7.11	&	95.43 	& C2\\
10	&	05	47	5.7	&	21	00	30.2	&	9.02	$\pm$	2.99	&	$-$45.97	$\pm$	4.98	&	44.03 &	 C1\\
11	&	05	47	4.3	&	21	00	09.6    &	11.37	$\pm$	4.54	&	19.92	$\pm$	10.59	&	109.92 &	 C2\\
12	&	05	47	2.9	&	21	00	10.2	&	13.98	$\pm$	4.36	&	$-$56.38	$\pm$	10.68	&	33.62 &	 C2\\
13	&	05	47	2.1	&	20	59	50.4	&	10.77	$\pm$	5.08	&	35.59	$\pm$	19.64	&	125.59 &	 C2\\
14	&	05	47	2.1	&	21	00	29.4	&	12.61	$\pm$	5.36	&	$-$25.51	$\pm$	11.15	&	64.49 &	 C2\\
15	&	05	47	1.4	&	21	00	20.2	&	12.71	$\pm$	3.09	&	10.61	$\pm$	5.48	&	100.61 &	 C2\\
\hline
\end{tabular}
\label{5}
\end{table*}


\subsection{Relative orientation between magnetic field and the Galactic plane}
The mean value of position angle of polarization (optical) and standard deviation for observed field stars are $<\theta^{\textrm{opt}> \sim 143.2^{\circ}}$ and $\sigma_\theta \sim 7.8^{\circ}$. The position angle of Galactic plane at the latitude of CB 34 is $\theta_{GP} \sim 148.7^\circ$ and $|\theta_{GP} - <\theta^{\textrm{opt}}>| \sim 5.5^\circ$. This suggests that the magnetic field in the periphery of CB34 is well aligned with the Galactic magnetic field in the observed plane of the sky , i.e., magnetic field in the projected plane of CB34 appears to be coupled with the Galactic magnetic field. This type of orientation is mostly pronounced at low galactic latitude like that of CB34 ($b = -3.83^\circ$). It is observed that the magnetic field within few hundred parsec of the Sun largely lies in the plane of the Galaxy \citep{Ma1970}. However, \cite{St2011} studied the effect of the Milky Way's magnetic field in star-forming regions using archival 350$\mu m$ polarization data on 52 Galactic star formation regions from the Hertz polarimeter module. They found that there is no correlation between mean polarization angle and Galactic location, i.e., the magnetic field in dense molecular clouds is decoupled from the large-scale Galactic magnetic field. The authors also commented that the cloud cores which are embedded in a diffuse medium (in an ordered Galactic magnetic field) usually have a meaningful net field that has no preferred direction within the Galaxy.
 It appears from our study that the magnetic field in the immediate environment (at scales $> 10^5$AU) is dominated by the galactic magnetic field. But, the submillimeter polarization measurements show that the magnetic field direction in the dense regions of the globule is different from the galactic magnetic field direction.

However, the minor axes of the two cores (C1 and C2) (please see Table \ref {7}) do not show any relation to either of these directions. So, it is evident that the processes leading to the formation of the globule CB34 has not affected the ambient magnetic field to any large extent.  This globule did not show pinching of the field lines or other signs that the field has been compressed due to collapse. The mean polarization angle of all field stars, then, most likely probe the average magnetic field direction of the general ISM along the line of sight to the star, and not the immediate environment of the globule.

To investigate the direction of  observed polarization vectors in the peripheral region with intercloud polarization vectors, we have obtained stellar polarization data of stars from \citet{He2000} within a circular area of radius 5$^{\circ}$ about the central coordinates of CB34. The total number of stars in that circular region is 24 which have $p/E_p > 3$. The mean and the standard deviation of polarization and position angle of polarization are estimated to be $2.46 \pm 0.04$\% and $146 \pm 40^\circ$. It further confirms that the direction of peripheral magnetic field in CB34 is also well aligned with the intercloud magnetic field.

The offset angle between the Galactic plane and magnetic field of two cores (C1 and C2) are found to be 102$^{\circ}$ and 58.3$^{\circ}$. This suggests that the inner magnetic field of CB34 is not coupled with the galactic magnetic field.

\section{Mean particle density}

 The density structure of a star forming cloud cores is one of the most important physical quantities that explores the evolution of the protostellar collapse and its stellar end product.
The mean particle density $<n_{H_2}>$ is given by \citep{Pe2007}
\begin{equation}\label{2}
    <n_{H_2}> = <A_V> \left(\frac{N_{H_2}}{A_V}\right)\frac{1}{l}~~~~ cm^{-3},
\end{equation}
assuming it to be a cylindrical filament. Here, $A_V$ is the extinction, $N_{H_2}$ is the hydrogen column density and $l$ (in cm) is the typical dimension of the cloud.

The standard gas-to-extinction ratio is given by \citep{Bo78}
\begin{equation}\label{3}
   \left(\frac{N_{H_2}}{A_V}\right) = 0.94 \times 10^{21} ~~~~cm^{-2}mag^{-1},
\end{equation}

We will now derive the mean particle density in both the high density and low density region.

\subsection{High density region}
 \cite{Wa1995} reported $N_{H_2}$ value for globule CB34 from $C^{18}O ~J = 2\rightarrow 1$ study, which is given by $9.9\times 10^{21}$cm$^{-2}$. In \emph{Section 3.2}, the dimension of the two cores C1 and C2 have been estimated and is given by $\sim 36''$ and $\sim 35''$. Using these data, it is possible to estimate the mean particle density at the central region of the cloud, especially at cores C1 and C2. The values are given by $1.22 \times 10^4$ cm$^{-3}$
 and $1.26 \times 10^4$ cm$^{-3}$.

\subsection{Low density region}
Near-infrared extinction measurements can detect dust column densities in the low-density outer regions of the globules. To construct the extinction map of CB34, we have used the Near-Infrared Color Excess (NICE) method developed by \cite{Ro2009} (RF). This method was first introduced by \cite{La1994} which combines measurement of NIR color excess to directly estimate the extinction and to map the dust column density through a molecular cloud. In RF's method, the median color of all stars at each pixel position has to be determined. Then each median color map has to be converted into the respective color excess map so that one can determine the extinction/column density of material. We used this method earlier to construct extinction map of two globules CB224 and CB130 \citep{Ba2015,Ch2016}.

We have collected the $J$, $H$ and $K_s$ magnitudes of all field stars from the 2MASS Point Source Catalog in regions of $25' \times 25'$ centered on the globule CB34 \citep{Cu2003}, to build the extinction map. Only those stars are considered whose $JHK_s$ magnitude show photometric quality flag of ``AAA" which actually corresponds to SNR $>$ 10.

The infrared color excess is related to the visual extinction via extinction law (Rowles \& Froebrich 2009):

\begin{equation}
    A_V = \frac{5.689}{2}(A_{H,<J-H>} + A_{H,<H-K_s>}),
\end{equation}

where, \\
\\
$A_{H,<J-H>}=\frac{<J-H>}{(\frac{\lambda_{H}}{\lambda_{J}})^{\beta}-1}$
or
$A_{H,<H-K_s>}=\frac{<H-K_s>}{1-(\frac{\lambda_{K_s}}{\lambda_{H}})^{-\beta}}$
is the extinction in the H-band. Here, $<J-H>$ and $<H-K_s>$ are the color excess.

\begin{figure*}

\hspace{2cm}
\includegraphics[width=32pc, height=22pc]{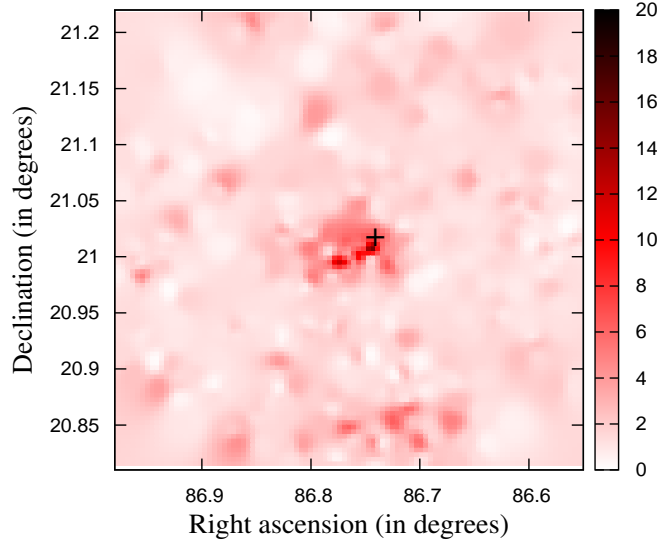}
\caption{Extinction map of CB34 constructed using the NICE method having field of view (FOV) $\sim 25' \times 25'$, dimension of a pixel $= 10 '' \times 10''$ (spatial resolution $= 34''$). The `+' sign denotes the center of the globule. The color bar represents the extinction scale which is in magnitude unit.}
\label{Fig4}
\end{figure*}

\cite{Ro2009} used the value of $\beta = 1.7$, since previous studies show that it lies between 1.6 and 2.0 \citep{Ma1990, Dr2003}. To derive extinction from equation (4), we have also taken $\beta = 1.7$. The visual extinction map of CB34 is now shown in Fig.~\ref{Fig4}. The field of view of visual extinction map is $\sim 25' \times 25'$, having dimension of each pixel $= 10 '' \times 10''$ (spatial resolution $= 34''$).

The  average extinction in the low density region can be estimated which is given by $\approx 1.80$ mag. The closest star which is traceable in the low density region is located at $\approx 100''$ from the center of globule. Thus we can determine the mean particle density using equation (2), considering the cloud dimension of $\approx 200''$, which is given by $3.76 \times 10^2$ cm$^{-3}$. The hydrogen column densities ($N_{H_2}$) for the region traced by optical observations can be calculated using equation (3) and is given by $1.7 \times 10^{21}$ cm$^{-2}$.

\section{Magnetic field strength}
We will now estimate the magnetic field strength (in Gauss) for SCUBA data using the relation \citep{Ch1953}:

\begin{equation}
    B = |B_{POS}| = \sqrt{\frac{4\pi}{3} \rho_{\textrm{gas}}} \,\,\,\, \frac{v_{\textrm{turb}}}{\sigma_{\theta}},
\end{equation}

where $\rho_{\textrm{gas}}$ is the gas density (in g cm$^{-3}$), $v_{\textrm{turb}}$ is the rms turbulence velocity (cm s$^{-1}$), and $\sigma_{\theta}$ is the standard deviation of the polarization position angles in radians. It is also assumed that the magnetic field is frozen in the cloud material.

The total gas density $\rho_{\textrm{gas}}$  is given by \citep{He1997}

\begin{equation}
\rho_{\textrm{gas}} = 1.36 \; n_{H_2} \; M_{H_2},
\end{equation}

where $M_{H_2}$ = 2.0158 amu = $2.0158 \times 1.66 \times 10^{-24}$ g is the mass of a H$_2$ molecule.

The rms turbulence velocity ($v_{\textrm{turb}}$) is given by \citep{Wa1995}

\begin{equation}
v_{\textrm{turb}} = \frac{\Delta v}{2.35},
\end{equation}

where $\Delta v$ is the FWHM line width, measured at quiescent positions located away from the emission peaks.

The mean density ($n_{H_2}$) of two cores and FWHM line width of CB34 are given by $1.2 \times 10^4 cm^{-3}$ and 1.5 $km s^{-1}$  \citep{Wa1995}. We have estimated the mean magnetic field strength of two cores C1 and C2 using the formulae mentioned above, where weighted standard deviation of position angle ($\sigma^{wt}_{\theta}$) of two cores is considered. The results are presented in Table \ref{6}. The magnetic field in core C2 is  $\approx 70\mu$G, which is more by a factor of 2, than the estimated field strength of core C1 $\approx 34\mu$G. It is observed that the sub-mm polarization vectors in core C1  are oriented more randomly than core C2. \citet{Co2003}, through multiline millimetre survey, reported that the current star forming activity in CB34 is concentrated in the three main clumps. They found that either CB34 is rotating, or that different parts of it are associated with different velocities. This may be the reason why the polarization vectors are not ordered at the core of CB 34.

\begin{table*}
 \begin{center}
  \caption{Position of cores, mean degree of polarization along with standard deviation,  intensity-weighted average magnetic field position angle along with weighted standard deviation and magnetic field strength of two sub-mm cores C1 and C2.}
  \begin{tabular}{cccccccc}
  \hline
Core  & RA (2000)  & DEC (2000)      & $<p_{\textrm{sub}}>$ & $\sigma_{p}$  & $<\theta^{sub}_{B}>_{wa}$ & $\sigma^{wt}_{\theta}$ & $B$  \\
        &         & & (\%) & (\%) &  ($^{\circ}$)  & ($^{\circ}$) & ($\mu$G)\\
\hline
C1 & 05 47 06 & 21 00 41 & 12.00 & 6.91 & $46.7$ & $51.9$ & $34$\\
C2 & 05 47 02 & 21 00 10 & 13.18 & 3.46 & $90.4$ & $25.5$ & $70$\\

\hline
\end{tabular}
\label{6}
\end{center}
\end{table*}

\begin{table*}
 \begin{center}
  \caption{Weighted average magnetic field position angles and angular orientations between magnetic field, core minor axis and outflows of two sub-mm cores C1 and C2 ($wa$: intensity-weighted average; $sub$: submillimetre data; $min$: minor axis of the core; $out$: outflow direction).}
  \begin{tabular}{ccccccc}
  \hline
Core & $<\theta^{sub}_{B}>_{wa}$ & $\theta_{min}$ & $\theta_{out}$  & $|<\theta^{sub}_{B}>_{wa} - \theta_{min}|$ &  $|<\theta^{sub}_{B}>_{wa} - \theta_{out}|$ & $|\theta_{out} - \theta_{min}|$ \\

 \hline
C1 & $46.7^{\circ}$  & 61$^{\circ}$ & $-15^{\circ}$&   $14.3^{\circ}$ & $61.7^{\circ}$ & 76$^{\circ}$ \\
C2 & $90.4^{\circ}$  & $-1^{\circ}$ & $-15^{\circ}$ &  $91.4^{\circ}$ & $105.4^{\circ}$ & 14$^{\circ}$ \\

\hline
\end{tabular}
\label{7}
\end{center}
\end{table*}

\begin{figure*}

\hspace{2cm}
\includegraphics[width=13cm, height=10cm]{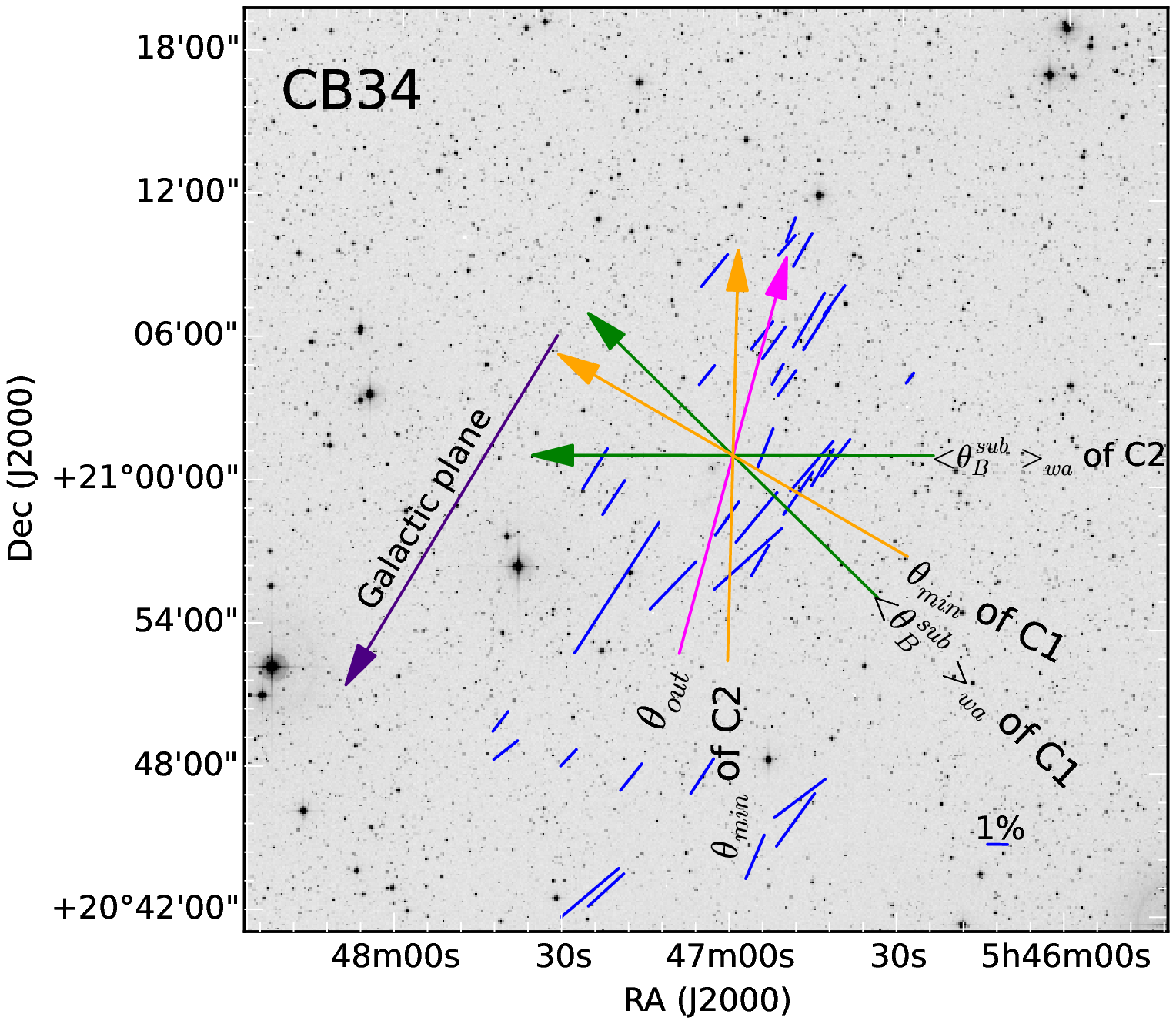}
\caption{Schematic diagram of CB34 showing the orientations of the intensity-weighted average magnetic field of two cores C1 and C2,  minor axis ($\theta_{min}$) of C1 and C2, direction of outflow ($\theta_{out}$), and the Galactic plane w.r.t. the north. Optical polarization vectors (blue lines) are overplotted on a large field $\sim 40' \times 40'$ R-band DSS image of the field containing CB34. A vector with a polarization of 1\% is drawn for reference which represents that the length of the line segments are proportional to polarization percentage.}
\label{Fig5}
\end{figure*}



\section{Discussion}
\subsection{Correlation between magnetic field and outflows}
Magnetic fields in a cloud play a notable role in the collapse of cloud and the formation of circumstellar disks and outflows. \citet{Ma2004} and \citet{Ma2006} showed that the alignment between the magnetic field and outflow depends on the magnetic field strength, i.e., the alignment will be better when the magnetic field is stronger.  This result is based on MHD simulations on a slowly rotating globule core which is undergoing gravitational collapse.
\cite{Cu2007} found no relation between mean magnetic field direction and outflow direction from the study of 16 high-mass star-forming regions, although some alignments were noticed. \cite{Hu2014} studied the correlation of B-fields with bipolar outflows and found that the sources with low polarization fractions show hint that outflows are preferentially perpendicular to small-scale B-fields. They commented that in those sources the fields have been wrapped up toroidally by envelope rotation.  Many observations have been reported to study the correlation between magnetic field and CO outflow of the Bok globules (e.g., \cite{Co1984,Vr1986,Jo2003,Wo2003,Hu2013,Be2014,So2015}). Some study shows a good alignment of globule's magnetic field with outflow axis (e.g., \cite{Co1984,Vr1986,Jo2003}), whereas misalignment of magnetic field direction and outflow axis have been also reported (e.g. \cite{Wo2003,Me2004,Hu2013,Ta2011,So2015}). \citet{So2015} reported that in IRAM 04191, the outflow is oriented almost perpendicular ($\sim 84^\circ$) to the peripheral magnetic field, but it is oriented almost parallel ($\sim 16^\circ$) to the inner magnetic field. In the case of globules B335, CB230 and CB68, the outflows are oriented almost perpendicular to the inner magnetic field \citep{Wo2003,Be2014}. \citet{Be2014} also reported that the magnetic field orientation of CB54 is aligned with CO outflow.

 The position angle of the bipolar outflow ($\theta_{out}$) of CB34 is $-15^\circ$ whose positions of the center of the outflows $\Delta \alpha$ and $\Delta \delta$ are given by $0.2'$ and $-0.4'$ (offset from the map center of CB34) \citep{Yu1994a}. Looking at the position coordinates of cores and outflow, we notice that the outflow center is very close to core C2. The angular offset between $<\theta^{sub}_{B}>_{wa}$ and $\theta_{out}$ for cores C1 and C2 are respectively, $61.7^\circ$ and $105.4^\circ$ (c.f Table \ref{7}).  Thus the outflow axis  is almost perpendicular to the magnetic field orientation of core C2. Orientation of both the polarization vectors and the outflow are shown in Fig.~\ref{Fig5}.

\subsection{Correlation between magnetic field, minor axis of the cores, and outflows}
We have estimated the orientation of the minor axis of two cores C1 and C2 which are given by $61^\circ$ and $-1^\circ$. It can be seen from Table \ref{7} that the angular offset between $<\theta^{sub}_B>_{wa}$  and $\theta_{min}$ for C1 and C2 are $14.3^\circ$ and $91.4^\circ$. Thus minor axis of the core C1 is almost aligned with inner magnetic field. This result is consistent with magnetically dominated star formation model which suggests that the magnetic field should lie along the minor axis of the molecular cloud \citep{mou1991,li98}. It is also expected that the cloud tends to contract first in a direction parallel to the magnetic field and then in quasi-statically perpendicular to the field orientation \citep{li98,Wa2009}. However, core C2 shows almost opposite nature where projected magnetic field vector is found to be perpendicular to the position angle of the minor axis, which is inconsistent with the above model. The mean value of offset between $\theta_{out}$ and $\theta_{min}$ for C1 and C2 are found to be $76^\circ$ and $14^\circ$. This finding suggests that the the CO outflow is oriented almost parallel to the minor axis of C2. It is also to be noted that the outflow center is close to core C2 (which is already discussed in Section 5.1), so we believe this outflow may be associated with core C2. It is important to mention here is that the major and minor axes of cores C1 and C2 don't differ very much, and that the reader should take that into account when considering their results regarding magnetically-dominated versus not-magnetically-dominated star formation.

\citet{So2015} found that the angular offset between $\theta^{sub}_B$  and $\theta_{rot}$ for IRAM 04191 is 14$^\circ$, i.e., inner magnetic field is almost parallel to the minor axis of the globule. In the case of IRAM 04191 and L1521F, the outflows are oriented parallel to the minor axis of the clouds. Recently, \citet{Ch2013} reported the 350$\micron$ polarization observations of four low-mass cores containing Class 0 protostars: L483, L1157, L1448-IRS2 and Serp-FIR1. A strong correlation between minor axes and outflow direction was observed. They concluded that the outflow inclination angle  could be used as a proxy for the pseudodisk symmetry (minor) axis inclination angle.

\section{Conclusions}
\begin{enumerate}
 \item We present the optical imaging polarimetric observations of a large globule CB34 which were carried out
using the Aryabhatta Research Institute of observational sciencES (ARIES) Imaging Polarimeter mounted on Cassegrain
focus of the 1.04m Sampurnanand Telescope of ARIES, Nainital, in R photometric band, on 12--13 March, 2013 and on 20 Feb, 2015.
 The mean value of  optical polarization and position angle for
    stars  projected in CB34  are $<p^{\textrm{opt}}>  = 2.14
    \%$ and  $<\theta^{\textrm{opt}}>$ = 143.2$^{\circ}$ with
    a   standard   deviation    of   $\sigma_{p}$   =   0.84$\%$   and
    $\sigma_{\theta}$  =  7.8$^{\circ}$,  respectively.

      The mean polarization ($<p^{\textrm{sub}}>$) and mean position angle ($<\theta^{sub}>$) of core C1 along with standard deviations are given by $12.0 \pm 6.91\%$  and $-24.63 \pm 69.30^\circ$ (or $155.37 \pm 69.30^\circ$). The values for C2 are given by $13.18 \pm 3.46\%$  and $-1.72 \pm 33.52^\circ$ (or $178.28 \pm 33.52^\circ$). The intensity-weighted average position angle ($<\theta^{sub}>_{wa}$) for cores C1 and C2 are given by $-43.3\pm6.5^\circ$ and $0.4\pm12.3^\circ$. Also, the weighted standard deviation of position angle ($\sigma^{wt}_{\theta}$) for cores C1 and C2 are estimated to be 51.9$^\circ$ and 25.5$^\circ$.

 \item  The direction of the magnetic field inside the dense region (traced through sub-mm polarimetry) is oriented approximately perpendicular to the magnetic field direction in the less dense outer regions of the globule. This implies that only the magnetic field in the dense regions is related to the physical processes inside the globule, while the magnetic field in the low-density region in the environment of the globule (traced through optical polarimetry) is not. The transition between the two regimes is around $10^5$ AU.

  \item The magnetic  field of core C2 is found to be almost perpendicular with
  the outflow direction of CB34 which is  also observed in  past for three clouds B335, CB230 and CB68 \citep{Wo2003,Be2014}. The  magnetic field strength in the  plane-of-sky for two cores C1 and C2 is estimated to be $B_{\textrm{sub}}$   $\approx  34\mu$G   and  $\approx 70\mu$G.

  \item The  angular offset between intensity-weighted average magnetic field position angle of core C1 and the minor   axis   position   angle  is   found   to   be $14.3^\circ$. This suggests that the magnetic field of core C1 is nearly aligned with the minor axis of the core. This feature is typical for magnetically dominated star formation models. We also find that the inner magnetic field of C2 is inclined at an angle  of $91.4^\circ$ with the minor  axis of the  cloud that is inconsistent with magnetically dominated star formation models. Since the major and minor axes of cores C1 and C2 don't differ very much, so one should take that into account when considering his results regarding magnetically-dominated versus not-magnetically-dominated star formation.

  \item The mean  value of offset between the minor  axis of the core C2
    and the  outflow directions  is found to  be $14^\circ$.  Thus the
    direction of outflow is almost  aligned with the minor axis of the
    core C2.
\end{enumerate}

\section*{Acknowledgements}
We are thankful to ARIES, Nainital for providing us the Telescope time. This work makes use of data products from the CADC repository of the SCUBA Polarimeter Legacy Catalogue and is highly acknowledged. We also acknowledge the use of the VizieR database of astronomical catalogues namely Two Micron All Sky Survey (2MASS), which is a joint project of the University of Massachusetts and the Infrared Processing and Analysis Center/California Institute of Technology, funded by the National Aeronautics and Space Administration and the National Science Foundation. We are thankful to anonymous referee for helpful suggestions which definitely helped to improve the quality of the paper. This work is supported by the Science and Engineering Research Board (SERB), a statutory body under Department of Science and Technology (DST), Government of India, under Fast Track scheme for Young Scientist (SR/FTP/PS-092/2011).







\end{document}